\documentclass[11pt,twoside]{article}

\usepackage{asp2004}
\usepackage{epsf}
\usepackage{psfig}
\usepackage{lscape}

\usepackage{textcomp}
\begin{document}
\title{The 24\textmu m Luminosity Function of spectroscopic SWIRE sources from the Lockman Validation Field.}
\author{N.Onyett$^1$, S.Oliver$^1$, G.Morrison$^4$, F.Owen$^2$, F.Pozzi$^3$, D.Carson$^1$, SWIRE team$^4$}   
\affil{$^1$Astronomy Centre, University of Sussex, Brighton, UK
$^2$NRAO, Socorro, USA
$^3$Dipartimento Astronomia Bologna, Italy
$^4$IPAC, CalTech, Pasadena, CA, USA}

\begin{abstract}A spectroscopic follow-up of SWIRE sources from the
  Lockman Validation Field has allowed the determination of the
  SWIRE/WIYN 24\textmu m Luminosity Function (LF). The spectroscopic
  sample was chosen above a 24\textmu m flux limit at 260\textmu Jy and an
  r-band optical limit of $r<21$. A spectroscopic completeness of
  82.5\% was achieved. We found the median redshift for the sample to
  be $z_{\rm med}=0.29$. Markov-Chain Monte-Carlo (MCMC) and Maximum
  Likelihood Estimator (MLE) techniques were employed to fit a
  parametric LF. Our result of the local LF (LLF) is consistent with
  the local 25\textmu m determination from Shupe et al .(1998). We
  split the sample at a redshift of $z_{\rm split}=0.36$ and find strong evidence for galaxy evolution.  
\end{abstract}

\section{Introduction}Since one can not study directly how a galaxy
  forms and evolves (how the mass and luminosity of a galaxy change
  with time), it becomes necessary to analyse the statistics of a
  large number of galaxies at different epochs. Using the SWIRE (Lonsdale et
  al. 2004) 24\textmu m MIPS-band, one is able observe the FIR
  emission from dust, tracing star-formation (SF) regions and
  SF-galaxies. By studying these galaxies at different epochs one can
  gain an estimate of the star formation history of the Universe.
   As a follow up to SWIRE, a spectroscopic redshift survey
  covering a 0.55 deg$^2$ region, centred in the Lockman Validation
  Field was undertaken. The sample was observed using the HYDRA
  spectrometer on the 4-m WIYN telescope at KPNO and included
  24\textmu m sources down to a Vega r-band magnitude of $r<21$. A
  total of 286 24\textmu m sources were observed and redshifts for the
  resultant spectra were determined. Redshifts for 170 sources with
  24\textmu m fluxes down to a flux limit of 260\textmu Jy were
  determined, giving a spectroscopic completeness for the sample of 82.5\%.

\section{Constructing SWIRE/WIYN LF$_{24}$}In order to
determine each object's 24\textmu m luminosity (L$_{24}$) an appropriate
SED and K-correction for that galaxy must be chosen. Here 6 SEDs were
chosen, 2 AGN, 2 spiral and 2 starburst (Xu et al., 2001). The
fitting was performed using a minimum $\chi^2$ technique that incorporated
an error weighting (including a dispersion factor, previously unseen in
the literature). This dispersion factor accounts for the scatter
within a particular population of galaxies in colour-colour
space. Each SED was then convolved through the 24\textmu m filter. The
LF$_{24}$ was estimated by implementing the 1/Vmax estimator technique
(Schmidt 1968) and took into account both the r-band and 24\textmu m
flux limits. A redshift split at $z_{\rm split}=0.36$ was imposed,
allowing evolution in the sample to be investigated, see Fig.1. The
high-z sample (pale line) is composed of 38\% of the total z-sample
with V/Vmax=0.72$\pm$0.04, showing strong evidence for luminosity
evolution. The low-z sample (dark line) is composed of 62\% of the
total z-sample with V/Vmax=0.46$\pm$0.03. Note: incompleteness corrections
need to be applied to the faintest luminosity bins of the high-z
sample.
\begin{figure}\plotfiddle{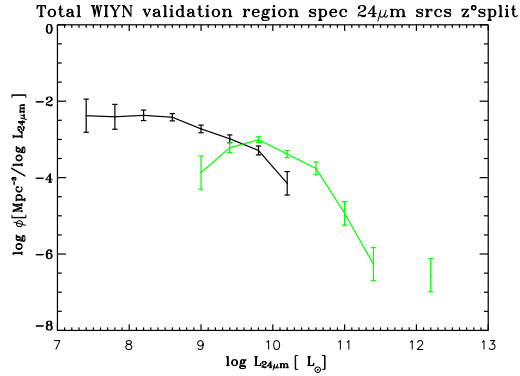}{3.5cm}{0}{40}{40}{-135}{-150}\caption{High [pale-line] \& low [dark-line] z-sample LF$_{24}$, spilt at z=0.36. Note that incompleteness corrections have not been made in faintest bins.}\end{figure}
 Parameterization of the SWIRE LF was performed and a comparison with
 low-z from IRAS was made. Using MCMC (Kashyap 1998) \& MLE (Sandage,
 et al. 1979) techniques to fit
 Saunders et al. 1990 parametric form of the LF, the best fitting
 MCMC and MLE model fits to the SWIRE LF$_{24}$ are shown in Fig.2. A
 direct comparison to Shupe et al. 1998 LF$_{25}$ is also shown in
 Fig.2. Since this is an apriori model and not a fit it provides a fairly good description of our data. 
\begin{figure}\plottwo{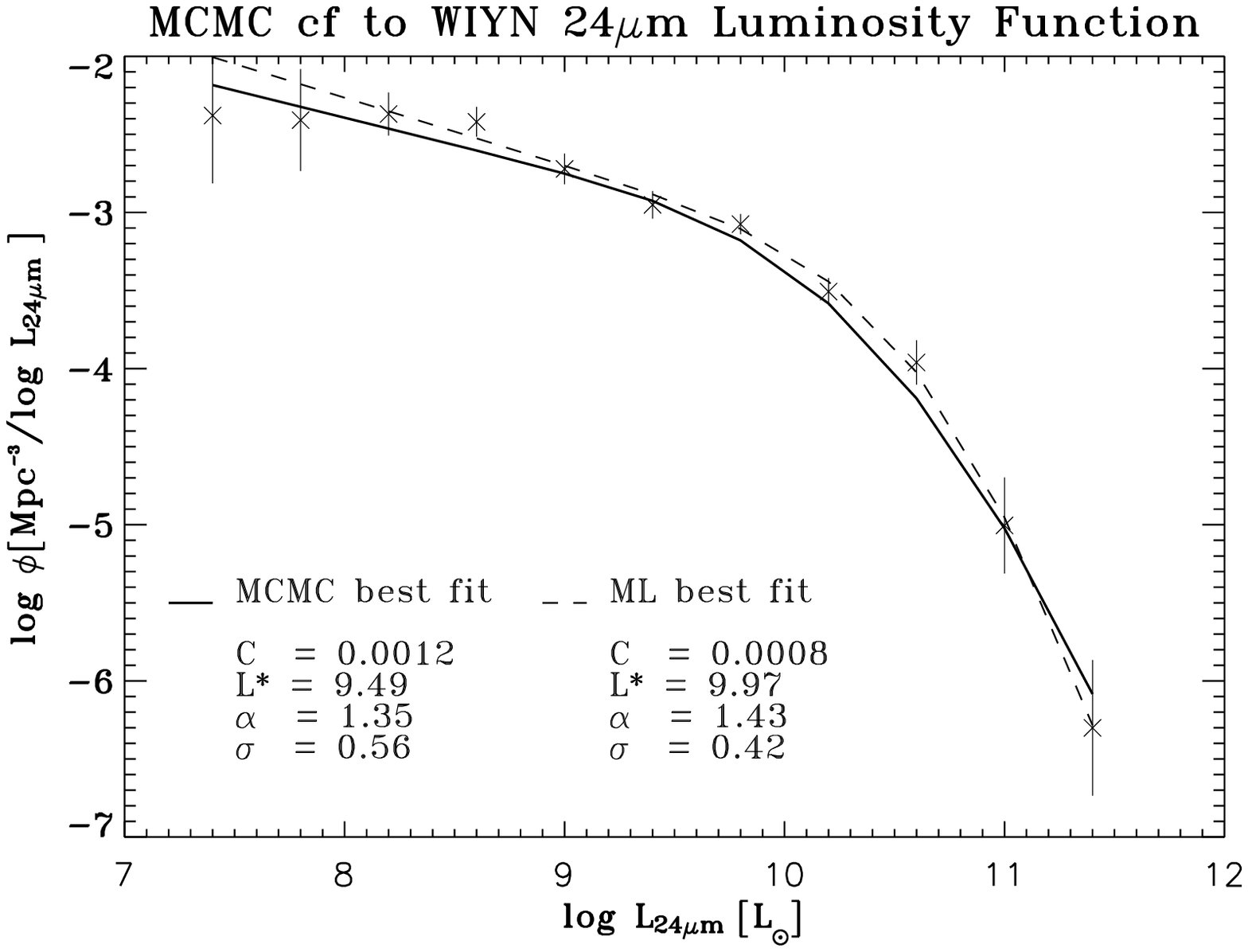}{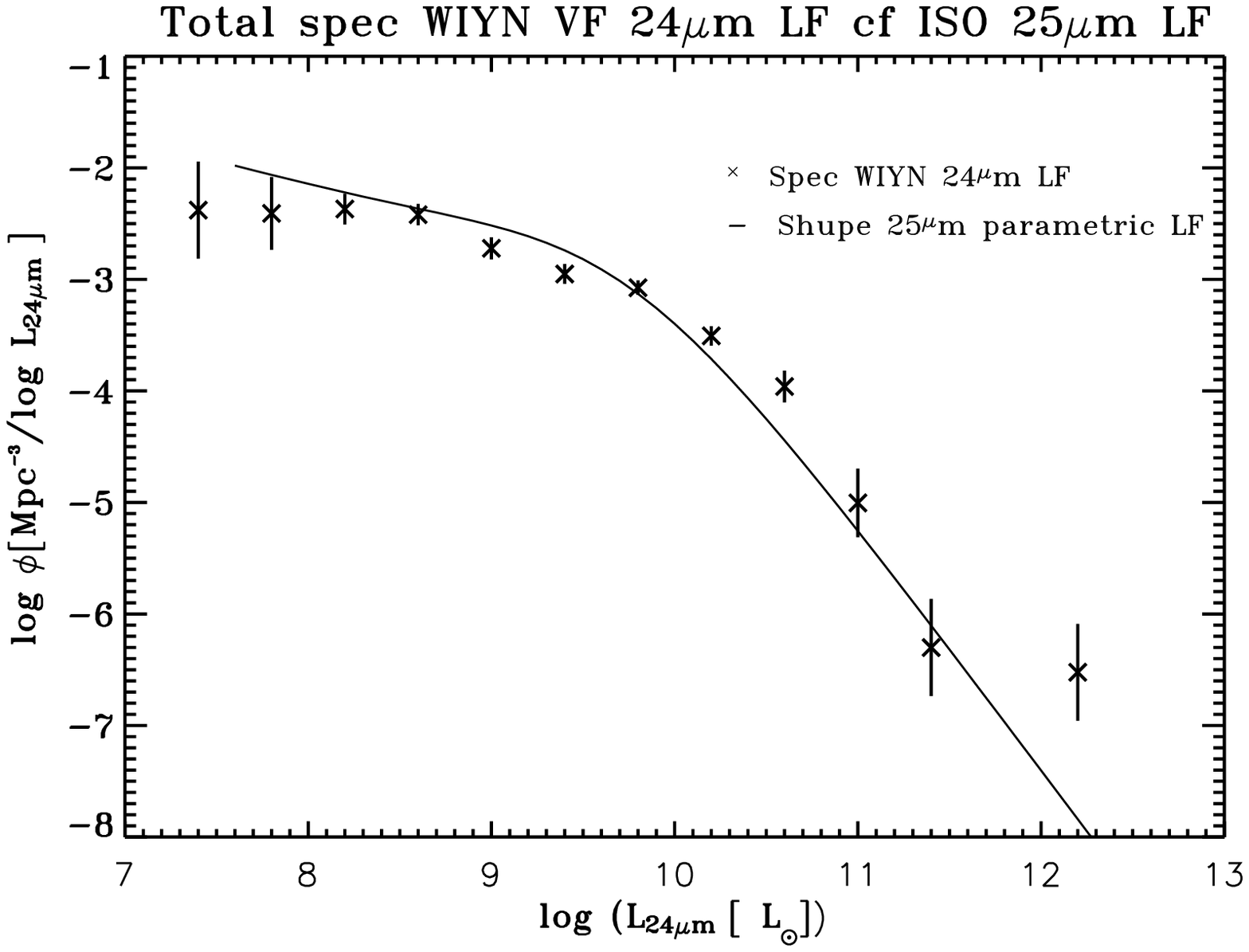}\caption{Left:MCMC
    \& MLE best fits Right:comparison to Shupe's LF$_{25}$}\end{figure}

\end{document}